\documentclass[draft]{agujournal2019}
\usepackage{url} 
\usepackage{lineno}
\usepackage{soul}
\draftfalse
\journalname{Space Weather}

\usepackage{graphicx}
\setkeys{Gin}{draft=false}
\usepackage{amssymb}
\usepackage{float}
\usepackage{rotating}
\usepackage{enumitem}
\usepackage{rotating}
\usepackage{makecell}
\usepackage{appendix}
\usepackage{fixmath}
\usepackage{epstopdf}
\usepackage{times} 
\usepackage{amsmath}
\usepackage{booktabs}
\usepackage{url}

\newcommand{\x}{\ensuremath{\times}}

\newcommand{\cmea}{CME 1}
\newcommand{\cmeb}{CME 3}
\newcommand{\cmec}{CME 11}
\newcommand{\cmed}{CME 16}
\newcommand{\lati}{$9.37 \pm 8.06\,^\circ$}
\newcommand{\longi}{$5.78 \pm 6.56\,^\circ$}
\newcommand{\atime}{$37.05 \pm 29.71\,\mathrm{minutes}$}
\newcommand{\dist}{$11.01 \pm 10.39\,^\circ$}
\newcommand{\fulllati}{$9.91 \pm 8.21\,^\circ$}
\newcommand{\fulllongi}{$16.66 \pm 49.08\,^\circ$}
\newcommand{\fullatime}{$38.45 \pm 29.59\,\mathrm{minutes}$}
\newcommand{\detections}{32}

\begin{document}
%
%

\correspondingauthor{T. Williams}{tomwilliamsphd@gmail.com}

%
%

\title{Automated detection of coronaL MAss ejecta origiNs for space weather AppliCations (ALMANAC)}
%

%
%

\authors{Thomas Williams\affil{1}, Huw Morgan\affil{1}}
\affiliation{1}{{Department of Physics, Aberystwyth University, Penglais, Aberystwyth, SY23 3BZ, UK}}

\begin{keypoints}
\item The software package presented can forecast the early signatures of coronal mass ejections (CMEs) in the low solar atmosphere in real time.
\item The goal of this work is to improve the forecasting of CME arrival times and potential impact when used as part of a software suite.
\item Applied to historical data\,-\,sets, the method can lead to a greater scientific understanding on the connection between CMEs and space weather.
\end{keypoints}

\begin{abstract}
Alerts of potentially hazardous coronal mass ejections (CME) are based on the detection of rapid changes in remote observations of the solar atmosphere. This paper presents a method that detects and estimates the central coordinates of CME eruptions in Extreme Ultraviolet (EUV) data, with the dual aim of providing an early alert, and giving an initial estimate of the CME direction of propagation to a CME geometrical model. In particular, we plan to link the ALMANAC method to the CME detection and characterisation module of the Space Weather Empirical Ensemble Package (SWEEP), which is a fully automated modular software package for operational space weather capability currently being developed for the UK Meteorological Office.  In this work, ALMANAC is applied to observations by the Atmospheric Imaging Assembly (AIA) aboard the Solar Dynamics Observatory (SDO). As well as presenting the method, a proof of concept test is made on a limited set of data associated with twenty halo CMEs recorded by the Coordinated Data Analysis Workshop (CDAW) catalogue near the activity maximum of solar cycle 24. SDO/AIA data for each event is processed at 6 minute cadence to identify the on-disk location and time of each CME. The absolute mean deviance between the ALMANAC and CDAW source event coordinates are within \atime\ and \dist. These promising results give a solid foundation for future work, and will provide initial constraints to an automated CME alert and forecasting system.
\end{abstract}

\section*{Plain Language Summary}
This work details a new tool to be used in the operational space weather forecasting of large eruptions from the solar atmosphere for the UK Meteorology Office. This work has considerable potential for economical and societal impact, since early warning of large space weather events can mitigate the substantial risk to technological infrastructure. We have created a novel software package which can forecast the early signatures of coronal mass ejections (CMEs) in the low solar atmosphere in real time. The code will eventually be used on a daily basis to improve predictions of CME arrival times and their potential impact at Earth. Furthermore, when the method is applied to historical datasets, it may lead to a greater scientific understanding on the connection between solar eruptive events and space weather.

\section{Introduction}\label{sec:intro}
Identifying and characterising coronal mass ejections (CME) in solar remote sensing observations is central to space weather forecasting. CMEs that are Earth-directed -- that is, halo\,-\,CMEs -- and with a large southerly-oriented $B_z$ component of the interplanetary magnetic field, can cause large disturbances in Dst, with potential large-scale disruption to technological infrastructure \cite{eastwood2018}.

Critical to providing early warning of potentially hazardous CMEs is the ability to estimate their main direction of propagation, and their true velocity, near the Sun. This requires constraints on the three-dimensional (3D) distribution of the CME material. Whilst current operational systems are based on manual fitting of CMEs observed by space-based coronagraphs from two viewpoints, GCS (Graduate cylindrical shell modelling \cite{thernisien09}) has been widely extended to three viewpoints \cite{jang16,mostl18,cremades20} and for use with ground\,-\,based data \cite{majumdar22}. The fitting is geometrical - that is, parameters describing a simple `wire-frame' model of a CME \cite{thernisien2006, zhao2002} are manually adjusted to visually match, or encompass, the CME activity in near real-time (NRT) time-differenced coronagraph images. Works using such geometrical fitting for scientific purposes include \citeA{temmer2012}, \citeA{hutton2015}, and \citeA{kane2021}.

Automated methods for constraining the 3D characteristics of CMEs in coronagraph data face large challenges. A more basic detection of CMEs, without 3D information, is comparatively more straightforward. Several catalogues, based on different manual and automated methods, keep records of onset time, source location, solar flare association, plane-of-sky (apparent) speed/acceleration, plane-of-sky mass, and so on. One such catalogue is the Coordinated Data Analysis Workshop \cite[CDAW]{yashiro04,gopalswamy08}, which manually identifies and records CME events. Within this comprehensive catalogue exists a sub-category specifically for halo CMEs. Here, the source location of halo CMEs are defined by the Solar Geophysical Data (SGD) listing of an associated H\,-\,$\alpha$ flare in heliographic coordinates. However, if flare information is unavailable from SGD, then the source information is obtained from inner coronal images \cite{gopalswamy07}. On the autonomous side there are a few catalogues that record CMEs, such as Automatic Recognition of Transient Events and Marseille Inventory from synoptic maps (\citeA[ARTEMIS]{artemis}; \citeA[ARTEMIS\,-\,II]{artemis2}), Solar Eruptive Events Detection System \cite[SEEDS]{seeds}, coronal image processing \cite[CORIMP]{corimp1,corimp2} CME catalogue, and Computer Aided CME Tracking \cite[CACTus]{berghmans02}. However, many of these catalogues require science\,-\,grade data that is not available in real time for forecasting. Whilst some catalogues, such as SEEDS, are capable of using real time or beacon data to varying results, the majority of catalogues are unsuitable for real time forecasting and are better suited to CME classification.

Utilising CDAW, \citeA{gopalswamy10} noted that halo-CMEs only accounted for $\approx3$\,\% of events detected by the Solar and Helisopheric Observatory \cite[SOHO]{soho} until the end of 2007. Despite this small portion of CMEs being full-halos, these are the most important subcategory of events due to their ability to cause geomagnetic storms or geoeffectiveness \cite{zhao03,kim05,yermolaev05,gopalswamy07} with frontsided full (partial) halos being 65\,\% (40\,\%) geoeffective. Furthermore, \citeA{gopalswamy10} note that the majority of automatic detection methods do not identify the majority of halo-CMEs and thus manual catalogues such as CDAW serve as important references in the process of improving automated CME detection algorithms.

\citeA{yashiro08} examined the distribution of CME properties between a manual (CDAW) and autonomous (CACTus) catalogue. They found that whilst both had good agreement in the CME properties when their $\mathrm{width} > 120$\,\textdegree, there is a significant discrepancy for narrow CMEs ($\mathrm{width} \leq 30$\,\textdegree). On an event\,-\,by\,-\,event basis they found that CDAW may have missed 1000\,--\,2300 narrow CMEs (particularly during solar maximum) between April 1997 and December 2006. However, during the same period CACTus had 3800 false positive detections, and it is also suggested by \citeA{yashiro08} that CACTus has difficulty detecting fast CMEs. These results highlight the difficulty of detecting CMEs within coronagraph data by both manual and autonomous methods.

A recent study \cite{alshehhi21} developed an unsupervised K\,-\,mean clustering method for classifying and detecting CMEs using pretrained convolutional neural networks (CNN) on difference images. Utilising SOHO's Large Angle and Spectrometric Coronagraph \cite[LASCO]{lasco} data, they are able to recover $\approx70\,\%$ of CMEs with a processing time of 1\,--\,4\,s for two sequence images. Adopting methods similar to CACTus and SEEDS, \citeA{Patel21} employ parabolic Hough transforms to detect off\,-\,disk eruptions in the low corona. They note that should an Extreme UltraViolet (EUV) instrument be situated $\pm90^{\circ}$ to the Sun\,-\,Earth line then their method would be capable of providing characteristics and kinematics of potential halo\,-\,CMEs. Similarly, other methods exist for predicting CMEs in EUV data, such as \citeA[coronal dimmings]{attrill10} and \citeA[recurrent neural networks (RNN)]{Liu20}.

Additionally, \citeA{bemporad14,patel18} present on-board autonomous CME detection algorithms for the Solar Orbiter METIS and ADITYA\,-\,L1 missions, respectively that are based on intensity and area thresholding in difference images. Such processing methods aboard coronagraphs may significantly reduce the telemetry some 85\,\% whilst ensuring a CME detection rate of 70\,\%. Such algorithms are necessary where telemetry must be reduced due to decreased communication bandwidth when spacecraft are in deep space, such as at Lagrangian L1 \cite{patel18}. Furthermore, these algorithms may also be used to generate triggers for space weather forecasting after specifying specific criteria in CME detection.

On the solar disc, similar autonomous methods have been developed for the identification and analysis of coronal Moreton waves. One such example is the Coronal Pulse Identification and Tracking Algorithm \cite[CorPITA]{long14}, which uses an intensity\,-\,profile technique to identify the propagating pulse, tracking it throughout its evolution before returning estimates of its kinematics. More specifically, CorPITA implements a percentage base difference image as the foundation of detection. This is achieved by calculating the difference between the current image with the selected base image, and then scaling the difference according to the base image. In a similar vein, the Automated Wave Analysis and Reduction \cite[AWARE]{ireland19} algorithm first adopts the running difference persistence images \cite{thompson16} to isolate bright, propagating wavefronts in the corona. Afterwards, AWARE determines the velocity, acceleration and distance travelled by the wavefront along the corona using the Random Sample Consensus (RANSAC) algorithm.

Currently, observational constraints in the distance range of $\sim$2-6\,$r_{\odot}$ are deficient. As such, many operational space weather forecasting models have adopted techniques such as data assimilation. That is, where solar wind observations at 1\,AU are used to update the inner\,-\,boundary conditions at 30\,$r_{\odot}$ \cite{lang21} of propagation models such as the Heliospheric Upwind eXtrapolation with time dependence \cite[HUXt]{owens20}. Other examples are using numerical models from lower in the solar atmosphere/corona to provide initial conditions \cite{reiss20}. Similarly, \citeA{gopalswamy17} present a model that generates a flux rope from eruption data, which may then be fed into global magnetohydrodynamics (MHD) models to provide realistic forecasts. However, they also note that the flux rope may be affected by CME deflection and rotation, something that may be mitigated with semi-analytic models \cite{kay15}. \citeA{sarkar20} have subsequently extended the FRED model to the INterplanetary Flux ROpe Simulator INFROS), which predicts the magnetic field vectors of a CME in the interplanetary medium using an observationally constrained analytical model.

The Space Weather Empirical Ensemble Package (SWEEP) is a fully automated modular software package for operational space weather capability currently being developed for the UK Meteorological Office. A key module for this package automatically detects and characterises CMEs, including an estimate of the CME's main direction of propagation based on coronagraph images from multiple spacecraft (elements of the module are based on the Automated CME Triangulation (ACT) approach of \citeA{hutton17}). The CME's geometrical and kinematic characteristics, with errors, are then fed into the HUXt model in order to provide an ensemble forecast at Earth -- however, it is worth noting that CMEs often get deflected after their appearance in low\,-\,coronal EUV images \cite[for example]{majumdar20}. The module's estimates of CME geometrical parameters can be improved through the provision of initial estimates of central longitude and latitudes, as well as other characteristics, which may be provided from EUV data using the method described here. This provides the motivation for developing a simple and efficient method that detects events in the Solar Dynamics Observatory's \cite[SDO]{sdo} Atmospheric Imaging Assembly \cite[AIA]{aia} data. The method is presented in \S\,\ref{sec:method}. As an initial proof of concept, the method is tested on a selected sample of Halo CME events from the CDAW catalogue with results presented in \S\,\ref{sec:res}. We use this catalogue since it is very widely used as a standard reference, and as discussed by \citeA{yashiro08}, the catalogue manually identifies CMEs within SOHO/LASCO which should ensure that the CME is real and not a false detection, i.e. any randomly selected event used to verify the method outlined in this manuscript will indeed be a CME of known origin. Secondly, CDAW has a subcategory for Halo CMEs (CDAW halo\,-\,CME catalogue: \url{https://cdaw.gsfc.nasa.gov/CME_list/halo/halo.html} that may also be accessed directly from within an SSWIDL environment. Section \ref{sec:conc} presents the authors' concluding remarks.

\section{The ALMANAC Code}\label{sec:method}
In this section, the procedures adopted by the Automated detection of coronaL MAss ejecta origiNs for space weather AppliCations (ALMANAC) code are outlined. Currently, ALMANAC is an Interactive Data Language (IDL; version 8.4) module that incorporates routines from the SolarSoft (SSWIDL; \url{www.mssl.ucl.ac.uk/surf/sswdoc/solarsoft}) and IDL Coyote (\url{www.idlcoyote.com}) libraries to detect the location and time of significant events in the low solar corona. Note that at this stage, the code does not inherently distinguish low coronal events as CMEs or not. For operational applications, our plan is to use ALMANAC in conjunction with a coronagraph CME module, thus the CME detections from coronagraphs can be used to filter appropriate events from ALMANAC. To detect low coronal events, ALMANAC is designed to analyse near real time (NRT) data from SDO/AIA. However, it may also be used to analyse historical synoptic or full-resolution SDO/AIA data for scientific or statistical purposes. In this work, the wavelength used for CME detection is 193\,\AA\ due to this being the channel with the best signal \cite{williams21,odwyer10}Whilst other SDO/AIA channels may be selected by the user, this may require additional tweaks to the algorithm and careful consideration of the resulting detection(s) is required due to the different temperature ranges of these channels and the potential that they may be linked to different physical processes.

For operational purposes ALMANAC may be called at set time intervals (e.g. hourly), or via some other trigger, such as from current coronagraph detection algorithms, upon which it will obtain the previous 8 hours of SDO/AIA NRT data at a user-specified cadence. The NRT data\,-\,cube is then processed to determine if any potentially significant events occur within that observation window. The key information such as time, location (in Stonyhurst, Carrington, and Heliographic coordinates), as well as the relevant portion of the data\,-\,cube and header files are all outputted to file should further analysis be required. 

\subsection{ALMANAC Image Processing}
\begin{figure*}
\centerline{\includegraphics[width=\textwidth]{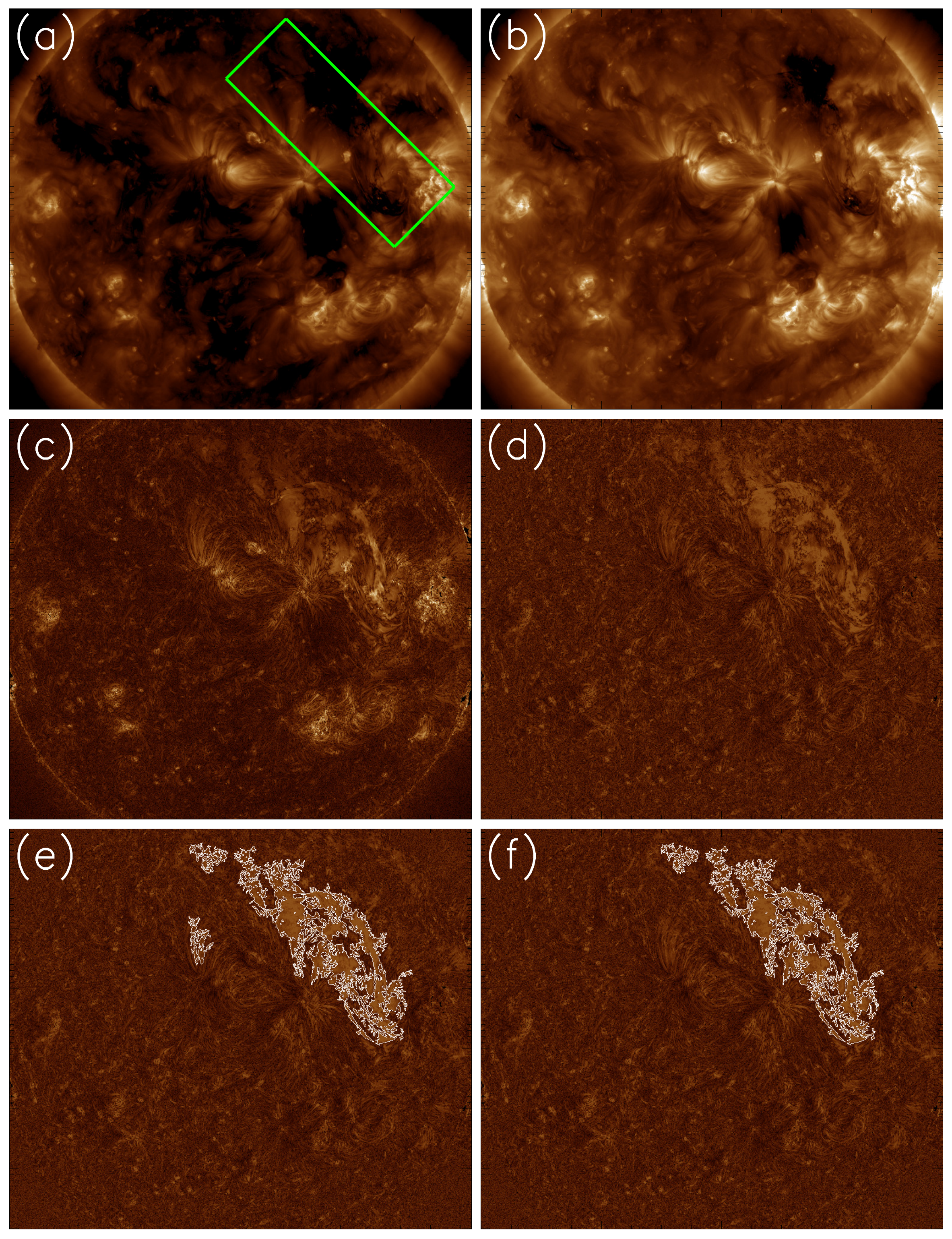}}
\caption{\textit{(a)}: Synoptic SDO/AIA image of the filament eruption (CME 12 in Table\,\ref{table:almanac}) with the source location boxed. The same image after normalisation is shown in (b) whilst (c) is after applying the high-bandpass filtering. Panel (d) shows a frame from `ratio' which is obtained by dividing (c) with the median time\,--\,averaged image. In (e) and (f) the Boolean mask are overlaid on-top of (d) before and after applying spatiotemporal thresholds, respectively.}
\label{fig:method}
\end{figure*}

To aid the description of how ALMANAC determines potential CMEs in the low corona, this subsection will discuss the detection process with reference to Figure\,\ref{fig:method}. This figure highlights the various steps within ALMANAC for a filament eruption, which is recorded as CME 12 in Table\,\ref{table:almanac}. Firstly, a spatiotemporal data\,-\,cube $[x,y,t]$ is loaded from individual FITS files using either full spatial resolution (4096\,$\times$\,4096\,px), synoptic (1024\,$\times$\,1024\,px), or NRT SDO/AIA data. For full spatial resolution images, ALMANAC rebins the data to spatial size 1024\,$\times$\,1024\,px. The spatial dimensions of the data\,-\,cube are then cropped to 75\% and 65\% in $x$ and $y$ respectively, which corresponds to an image size of 769\,$\times$\,667\,px. The cropping is centered on disk center such that the off-limb contribution and processing time are minimised (Figure\,\ref{fig:method}\,a). Intensities are thresholded to values between 0\,--\,2000\, DN\,s$^{-1}$. The data\,-\,cube is then normalised by dividing the data by its median. The data is then multiplied by a constant 150\,DN\,s$^{-1}$ (Figure\,\ref{fig:method}\,b). This trimming and normalisation method reduces flicker between images and provides a standardised intensity across each frame of the data\,-\,cube. 

The resulting trimmed and normalised data\,-\,cube, $D$, is smoothed over time through convolution with a kernel $k_t$ of width one hour: $D_s=D \otimes k_t$. This smoothed data\,-\,cube is subtracted from the non-smoothed data\,-\,cube to provide a high-bandpass time-filtered data\,-\,cube $D^\prime = D - D_s$ (Figure\,\ref{fig:method}\,c). Static, or slowly-changing intensities are effectively removed by this filtering and in combination with the initial thresholding and normalisation above, provides a stable data\,-\,cube that enables robust detection of rapidly-changing events. An image $M$ is then created as the median absolute values of $D^\prime$ for the 8\,-\,hour observation window. At each time step, the ratio $R_t=|D_t^\prime| / M$ is calculated (Figure\,\ref{fig:method}\,d) and stored as a new data\,-\,cube $R$.

The next step is to create a Boolean mask based on the ratio data\,-\,cube $R$, whereby regions are grouped and labelled as potential CMEs. For this, an initial mask is returned as true for values $R>2$ for connected clusters containing more than 1050\,pixels. This masking is done individually on each time step image in the data\,-\,cube, resulting in a mask data\,-\,cube $B$. This first step adopts a relatively lenient lower threshold on the spatial size of events such that moderately significant variations in the ratio data\,-\,cube are recorded. One issue that may occur when grouping regions is if the masked regions become disjointed they may be classed as independent events and thus labelled separately from each other. To mitigate this, the Boolean mask is converted to floating point numbers $0.0$ and $1.0$, and smoothed in the spatial dimensions by convolution with a narrow Gaussian kernel of width (standard deviation) $1 \times 1$ pixel. Following smoothing, a spatiotemporal Boolean mask is defined as true for values of the smoothed images greater than the median (approximately 0.2). This spatial smoothing helps join segmented regions and reduces the number of independent regions identified if they reside in close proximity to each other.

Another issue caused by the minimum thresholds set on the spatial coordinates is that some regions in the mask may disappear at time and reappear in a subsequent time step if their size momentarily decreases beneath the threshold. To combat this the mask is smoothed over time using a narrow Gaussian kernel of width 0.7 time steps, similar to the spatial smoothing described above.

Following these thresholding and smoothing procedures, we create a Boolean mask defined as regions of the smoothed mask that have values greater than 0.8 and containing 1500 or more connected pixels within a single frame. This masking is done individually on each time step image in the data\,-\,cube. Connected groups of voxels within this mask data\,-\,cube are then uniquely labelled. Since these regions are based on the smoothed mask, the labelled regions are then multiplied by the original mask $B$ (Figure\,\ref{fig:method}\,e). 

A final step is to discard connected regions that contain fewer than 9000 voxels, and that persist for a time period less than 18\,minutes (Figure\,\ref{fig:method}\,f). Regions that meet these criteria are our candidate events, from which we extract certain characteristics such as timing and location. The observation time of the first frame within the image sequence where the isolated region is identified is taken to be the onset time for that potential CME. Taking the first frame of the Boolean mask for each event, the central location of each event is defined by the center of mass (CoM) of the ratio $R$ for the pixels contained within the event region. The $x,y$ pixel numbers are then converted into longitude, latitude in the Stonyhurst and Carrington heliographic coordinate systems. These basic characteristics are sufficient for the results presented in this work, but we note that the masked regions, and the values at these pixels in the calibrated EUV images can be used for more advanced analysis.
\begin{table}
\centering
\caption{The results of CME source location obtained with ALMANAC for twenty example halo\,-\,CMEs identified from CDAW.}
\begin{tabular}{ |c|c|c|c|c|c|c|c| }
\hline
Index & Date & Time & X\,--\,ray & Flare Onset & Location & Time & Location \\
& & (LASCO C2) & (CDAW) & (CDAW) & (CDAW) & (ALMANAC) & (ALMANAC) \\
\hline
1 & 2010\,-\,08\,-\,14 & 10:12:05 & C\,4.4 & 09:38 & N\,17\,W\,52 & 08:40 & N\,9\,W\,61 \\
\hline
2 & 2011\,-\,02\,-\,15 & 02:24:05 & X\,2.2 & 01:44 & S\,20\,W\,12 & 00:30 & S\,17\,W\,5 \\
\hline
3 & 2011\,-\,09\,-\,06 & 02:24:05 & M\,5.3 & 01:35 & N\,14\,W\,7 & 01:20 & N\,13\,W\,10 \\
\hline
4 & 2011\,-\,11\,-\,26 & 07:12:16 & C\,1.2 & 06:39 & N\,17\,W\,49 & 06:20 & S\,12\,W\,46 \\
\hline
5 & 2012\,-\,04\,-\,07 & 21:15:59 & -\,-\,- & 19:25 & S\,24\,E\,168 & 18:20 & S\,4\,W\,55 \\
\hline
6 & 2012\,-\,11\,-\,23 & 13:48:06 & B\,5.8 & 11:00 & S\,38\,W\,10 & 12:20 & S\,43\,E\,18 \\
\hline
7 & 2012\,-\,11\,-\,27 & 02:36:05 & -\,-\,- & 00:09 & N\,13\,E\,68 & 01:30 & N\,32\,E\,83 \\
\hline
8 & 2013\,-\,03\,-\,15 & 07:12:05 & M\,1.1 & 05:46 & N\,11\,E\,12 & 05:50 & N\,5\,E\,6 \\
\hline
9 & 2013\,-\,04\,-\,11 & 07:24:06 & M\,6.5 & 06:55 & N\,9\,E\,12 & 06:40 & N\,3\,E\,15 \\
\hline
10 & 2013\,-\,05\,-\,15 & 07:12:05 & X\,1.2 & 01:25 & N\,12\,E\,64 & 01:00 & N\,17\,E\,59 \\
\hline
11 & 2013\,-\,07\,-\,09 & 15:12:09 & B7.7 & 14:00 & N\,19\,E\,14 & 14:00 & N\,23\,E\,18 \\
\hline
12 & 2013\,-\,08\,-\,20 & 08:12:05 & C\,1.1 & 06:16 & S\,31\,W\,18 & 06:40 & S\,38\,W\,20 \\
\hline
13 & 2013\,-\,09\,-\,29 & 22:12:05 & C\,1.3 & 21:43 & N\,17\,W\,29 & 20:50 & N\,31\,W\,29 \\
\hline
14 & 2014\,-\,02\,-\,18 & 01:36:21 & -\,-\,- & 00:30 & S\,24\,E\,34 & 00:30 & S\,38\,E\,42 \\
\hline
15 & 2014\,-\,03\,-\,23 & 03:36:05 & C\,5.0 & 03:05 & S\,12\,E\,40 & 02:20 & S\,13\,E\,31 \\
\hline
16 & 2014\,-\,04\,-\,01 & 16:48:05 & -\,-\,- & 14:00 & S\,9\,E\,12 & 15:10 & N\,3\,E\,12 \\
\hline
17 & 2014\,-\,04\,-\,29 & 23:24:05 & B\,9.1 & 22:28 & S\,12\,E\,15 & 22:30 & S\,9\,E\,17 \\
\hline
18 & 2014\,-\,08\,-\,15 & 17:48:07 & -\,-\,- & 16:16 & S\,10\,W\,5 & 16:50 & N\,8\,W\,7 \\
\hline
19 & 2014\,-\,08\,-\,22 & 11:12:05 & C\,2.2 & 10:13 & N\,12\,E\,1 & 09:50 & N\,12\,W\,0 \\
\hline
20 & 2014\,-\,12\,-\,21 & 12:12:05 & M\,1.0 & 12:12 & S\,14\,W\,25 & 10:50 & N\,8\,W\,26 \\
\hline

\end{tabular}
\label{table:almanac}
\end{table}
\begin{table}
\centering
\caption{ALMANAC Detections for random 8\,--\,hour periods where no events exist in the CDAW database.}
\begin{tabular}{ |c|c|c|c|c| }

\hline
Index & Start Time & End Time & ALMANAC & Event \\
& & & Detections & Type \\
\hline
1 & 2017\,-\,10\,-\,25  06:45 & 2017\,-\,10\,-\,25  14:45 & 0 &   \\
\hline
2 & 2018\,-\,01\,-\,31  15:00 & 2018\,-\,01\,-\,31  23:00 & 0 & \\
\hline
3 & 2018\,-\,06\,-\,26  17:00 & 2018\,-\,06\,-\,27  01:00 & 0 &   \\
\hline
4 & 2018\,-\,08\,-\,28  00:00 & 2018\,-\,08\,-\,28  08:00 & 1 & SDO/AIA jitter \\
\hline
5 & 2019\,-\,04\,-\,11  13:45 & 2019\,-\,04\,-\,11  21:45 & 2 & Evolving Loops \\
\hline
6 & 2019\,-\,05\,-\,25  06:55 & 2019\,-\,05\,-\,25  14:55 & 0 &   \\
\hline
7 & 2019\,-\,12\,-\,26  15:30 & 2019\,-\,12\,-\,26  23:30 & 0 &   \\
\hline
8 & 2020\,-\,03\,-\,12  20:50 & 2020\,-\,03\,-\,13  04:50 & 0 &   \\
\hline
9 & 2020\,-\,04\,-\,19  04:05 & 2020\,-\,04\,-\,19  12:05 & 0 &   \\
\hline
10 & 2020\,-\,04\,-\,22  21:25 & 2020\,-\,04\,-\,23  05:25 & 0 &   \\
\hline
11 & 2020\,-\,06\,-\,06  19:00 & 2020\,-\,06\,-\,07  03:00 & 0 &   \\
\hline
12 & 2020\,-\,07\,-\,03  12:05 & 2020\,-\,07\,-\,03  18:05 & 0 &   \\
\hline
13 & 2020\,-\,09\,-\,08  05:45 & 2020\,-\,09\,-\,08  13:45 & 0 &   \\
\hline
14 & 2021\,-\,01\,-\,09  18:05 & 2021\,-\,01\,-\,10  02:00 & 0 &   \\
\hline
15 & 2021\,-\,03\,-\,01  22:10 & 2021\,-\,03\,-\,02  06:10 & 0 &   \\
\hline
16 & 2021\,-\,05\,-\,13  09:20 & 2021\,-\,05\,-\,13  17:20 & 1 & CME \\
\hline
17 & 2021\,-\,06\,-\,28  15:35 & 2021\,-\,06\,-\,28  23:35 & 0 &   \\
\hline
18 & 2021\,-\,07\,-\,07  02:15 & 2021\,-\,07\,-\,07  10:15 & 0 &   \\
\hline
19 & 2021\,-\,10\,-\,31  11:30 & 2021\,-\,10\,-\,31  19:30 & 0 &  \\
\hline
20 & 2021\,-\,11\,-\,17  00:00 & 2021\,-\,11\,-\,17  08:00 & 0 &   \\
\hline

\end{tabular}
\label{table:random}
\end{table}
\begin{figure*}
\centerline{\includegraphics[width=0.7\textwidth]{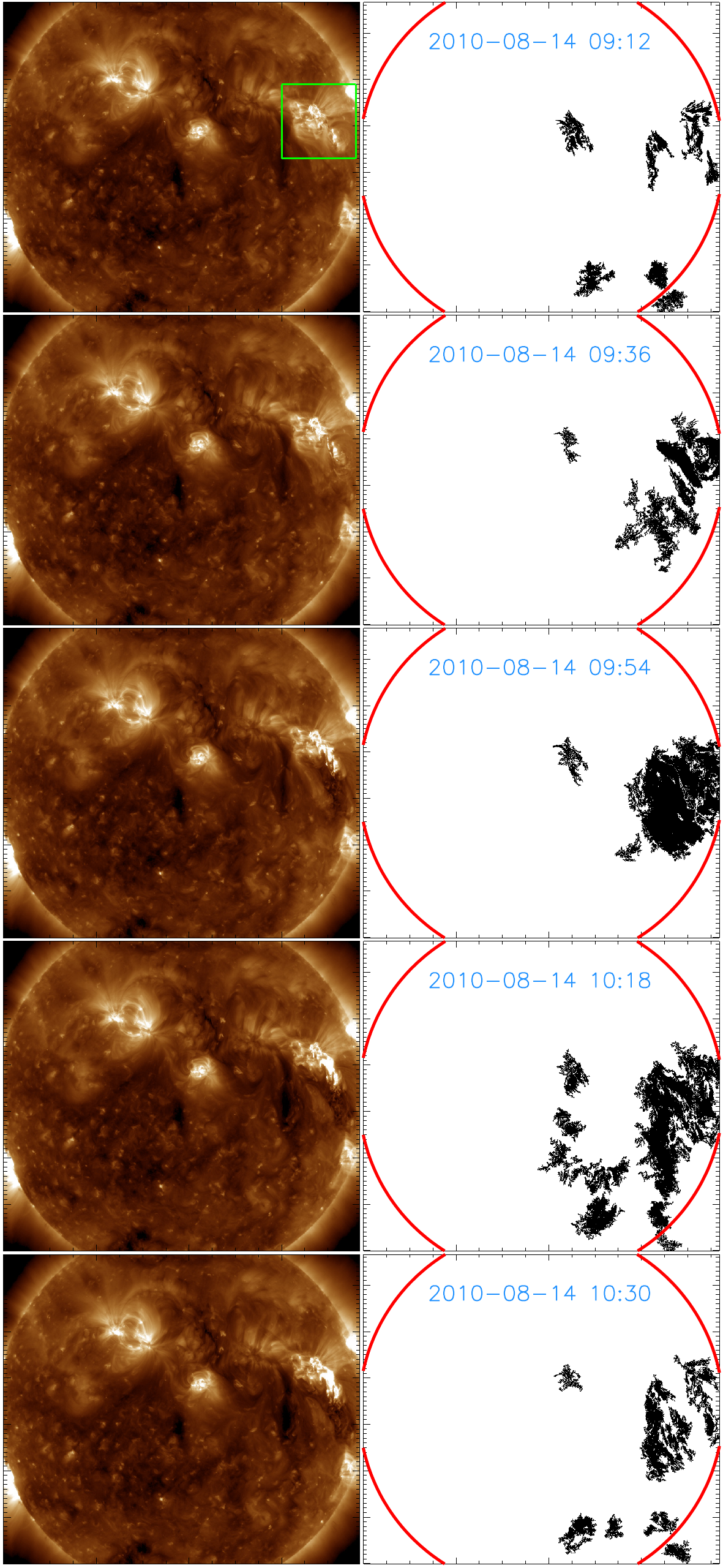}}
\caption{\textit{Left}: Synoptic SDO/AIA images taken at regular intervals (time increases in descending order) for \cmea. The green box in the first image indicates the region from which the CME emerges. \textit{Right}: The corresponding Boolean mask generated by ALMANAC isolating the eruptive pixels (black). The solar limb is indicated in red. Note that the eruption is difficult to see in the SDO/AIA images but is evident in the animated version of this figure (online only).}
\label{fig:example1}
\end{figure*}
\begin{figure*}
\centerline{\includegraphics[width=0.7\textwidth]{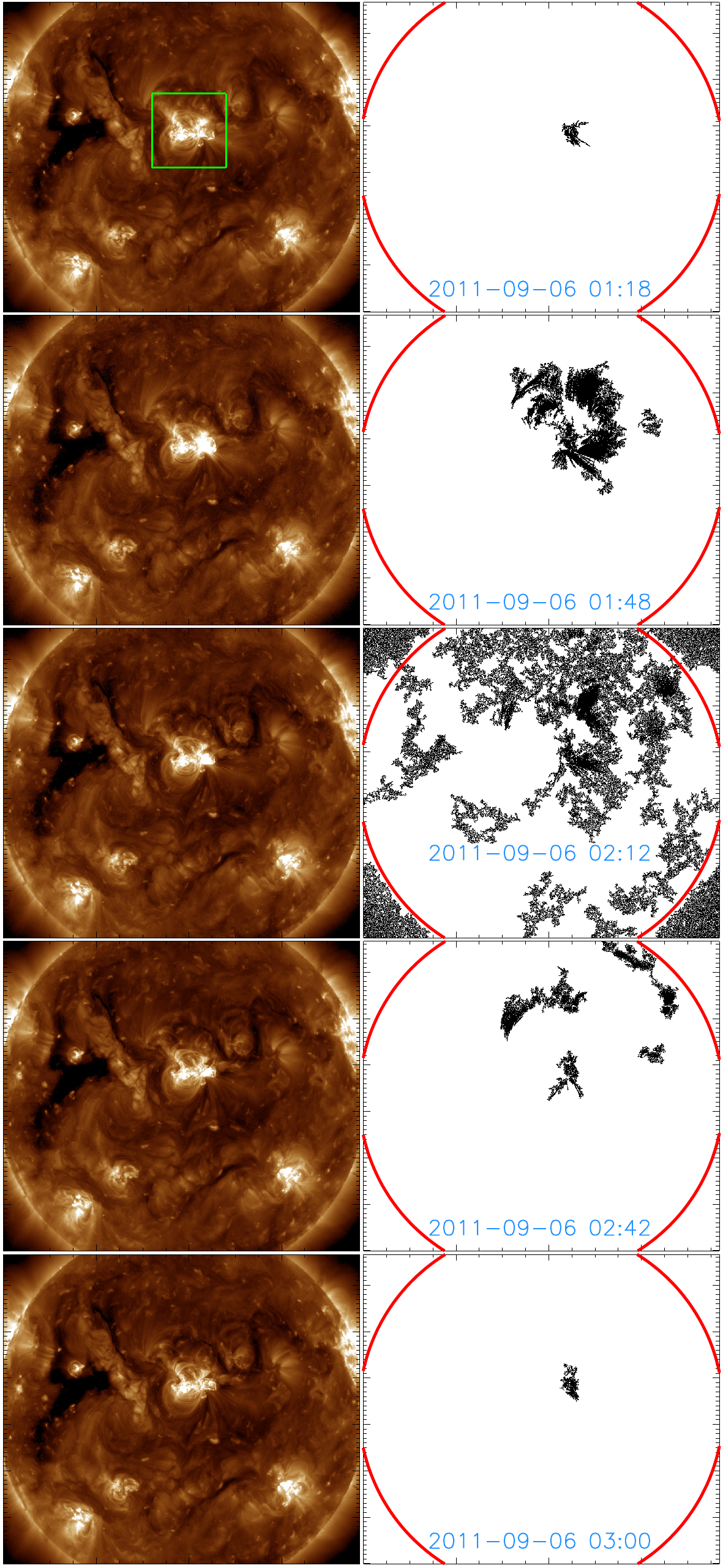}}
\caption{\textit{Left}: Synoptic SDO/AIA images taken at regular intervals (time increases in descending order) for \cmeb. The green box in the first image indicates the region from which the CME emerges. \textit{Right}: The corresponding Boolean mask generated by ALMANAC isolating the eruptive pixels (black). The solar limb is indicated in red. Note that the eruption is difficult to see in the SDO/AIA images but is evident in the animated version of this figure (online only).}
\label{fig:example2}
\end{figure*}
\begin{figure*}
\centerline{\includegraphics[width=0.7\textwidth]{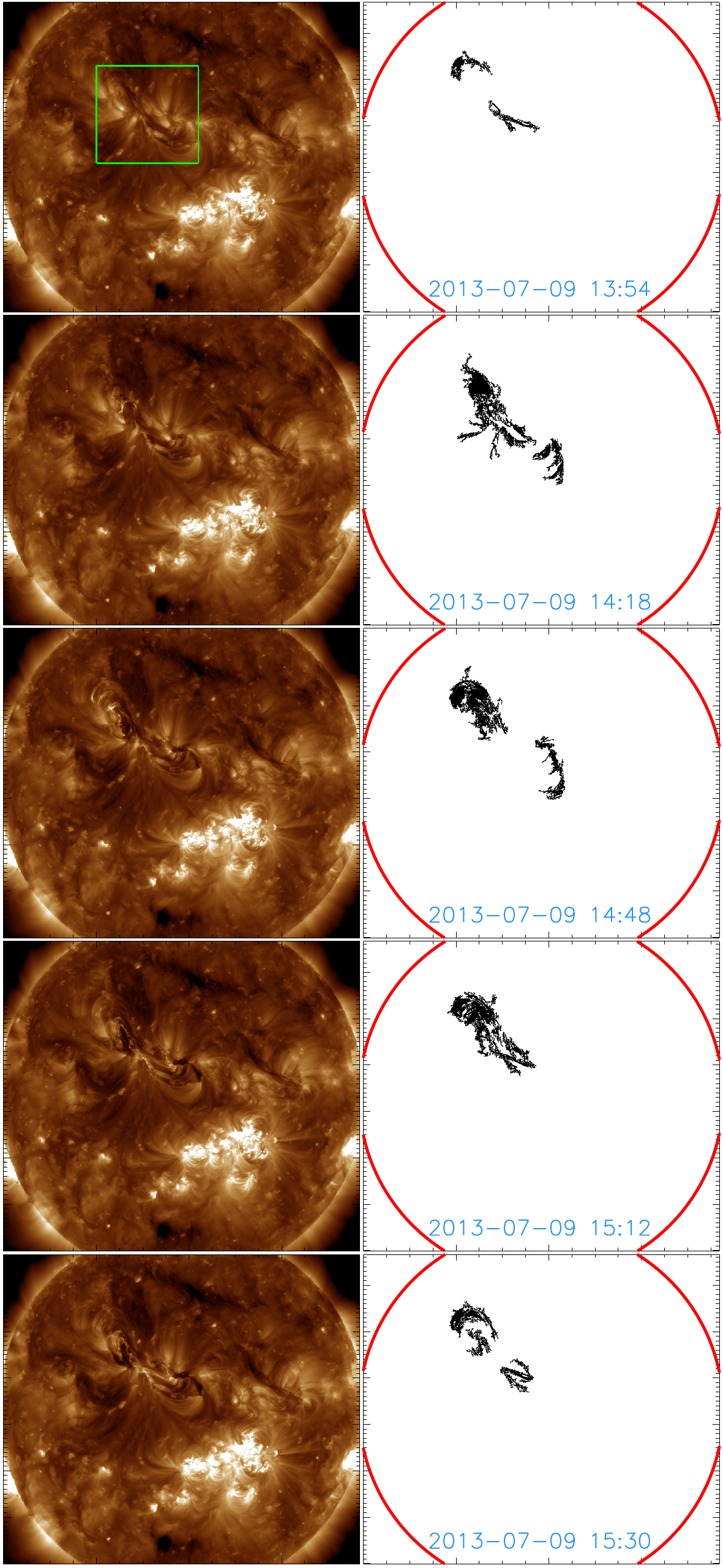}}
\caption{\textit{Left}: Synoptic SDO/AIA images taken at regular intervals (time increases in descending order) for \cmec. The green box in the first image indicates the region from which the CME emerges. \textit{Right}: The corresponding Boolean mask generated by ALMANAC isolating the eruptive pixels (black). The solar limb is indicated in red. Note that the eruption is difficult to see in the SDO/AIA images but is evident in the animated version of this figure (online only).}
\label{fig:example3}
\end{figure*}
\begin{figure*}
\centerline{\includegraphics[width=0.7\textwidth]{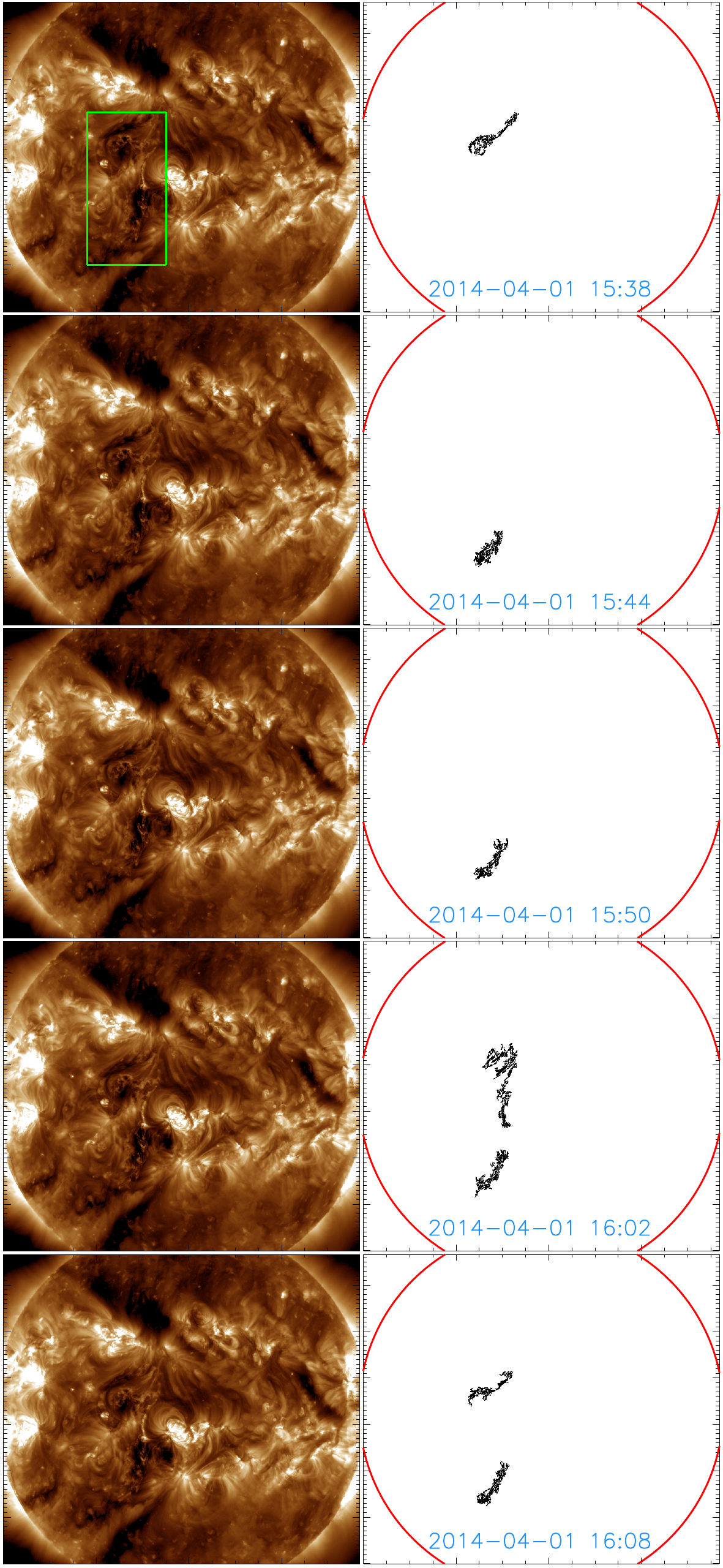}}
\caption{\textit{Left}: Synoptic SDO/AIA images taken at regular intervals (time increases in descending order) for \cmed. The green box in the first image indicates the region from which the CME emerges. \textit{Right}: The corresponding Boolean mask generated by ALMANAC isolating the eruptive pixels (black). The solar limb is indicated in red. Note that the eruption is difficult to see in the SDO/AIA images but is evident in the animated version of this figure (online only).}
\label{fig:example4}
\end{figure*}
\begin{figure*}
\centerline{\includegraphics[width=\textwidth]{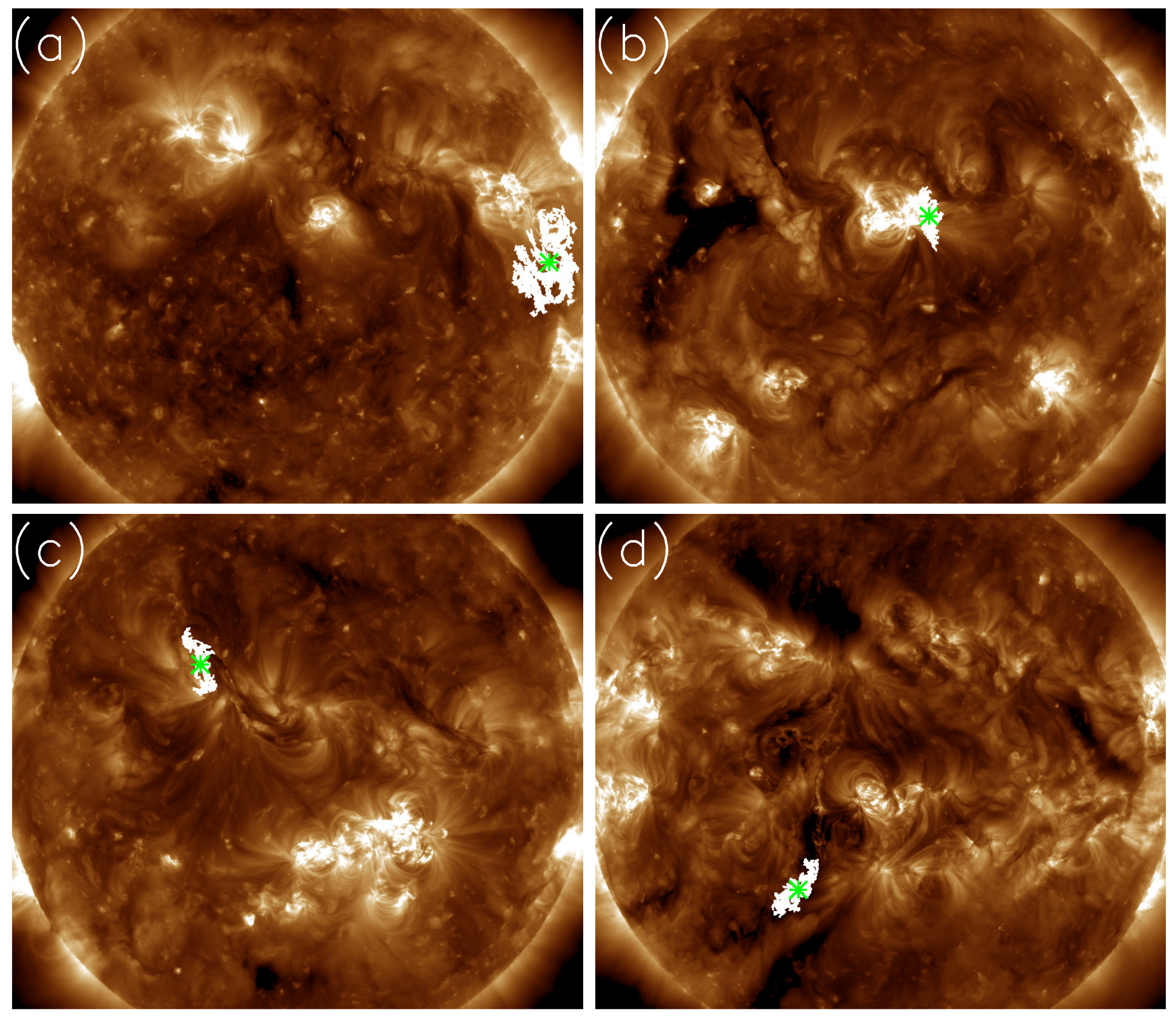}}
\caption{SDO/AIA images with the Boolean mask from ALMANAC overlaid on top (white). The centre of mass for each eruption is indicated by a green asterisk. Panels a\,--\,d show the erupting frame for CMEs 1 (a), 3 (b) , 11 (c), and 16 (d).}
\label{fig:com}
\end{figure*}

\section{Results and Analysis}\label{sec:res}
A total of twenty Halo CMEs are selected from the CDAW halo\,-\,CME catalogue (Table\,\ref{table:almanac}). Here, the reported Time and Location for ALMANAC indicate the first frame of detection and the centre of mass coordinates for each event. From these twenty halo\,-\,CMEs a total of 160\,hours of synoptic SDO/AIA 193\,\AA\ data are analysed where a total of \detections\ events are detected by ALMANAC. These events include the CMEs corresponding to the CDAW events as well as twelve other detections. These additional events recorded by ALMANAC are the result of bifurcation of connected regions, rapid changes in active region and/or loop topology (such as flares), and an additional halo CME that was present in the observation window (CDAW Halo CME 2011-02-14 18:24). 

To verify the robustness of ALMANAC, an additional twenty eight\,--\,hour periods (Table\,\ref{table:random}) are analysed where no events, halo\,-\,CME or otherwise, are listed in the CDAW catalogue. From these 160\,hours a total of four events are detected by ALMANAC. One of these are due to SDO/AIA `jitter' which may be caused by the instrument pointing suddenly changing or due to periods of missing data. Two events are associated with large-scale topological changes to evolving loops within an active region, and the remaining event is due to a near-side CME that was missed by the CDAW catalogue. Thus, ALMANAC is correctly and effectively detecting and recording temporal changes in the low corona, although it is prone to false positives. In particular, ALMANAC records large topological changes in coronal structures such as active region loops, which are not eruptive events.

\subsection{Testing ALMANAC against CDAW}
In this subsection, a comparison is made between the halo\,-\,CME source regions in CDAW and ALMANAC. For this, we have manually identified from the \detections\ ALMANAC events those that correspond to the 20 randomly selected CDAW halo CMEs between 2010 and 2014.

Focusing upon the individual events rather than the collective, the selection criteria for the CMEs in this study are as follows. Firstly, the CMEs selected occur at/near solar maximum as this will be the most difficult period to identify and isolate eruptions. Secondly, due to the fact that CMEs may be triggered in a number of ways, events with various corresponding soft X\,-\,ray emission must be sampled. This will allow ALMANAC to be tested for a variety of eruptions and not just one or two specific scenarios. As such, four examples are selected for closer consideration in this manuscript; CME indexes 1, 3, 11, and 16 (Table\,\ref{table:almanac}). These four examples include CMEs associated with C and M class flares along with two eruptions with no associated flares, allowing the CME detection ability of ALMANAC to be demonstrated across a variety of events.

The evolution of \cmea\ is shown in Figure\,\ref{fig:example1}. On the left are the synoptic SDO/AIA images shown over time from top to bottom. The first (top) frame is recorded at 09:12 and the last (bottom) frame at 10:30. The green box on the first SDO/AIA frame indicates the region from which the CME originates. On the right are the corresponding Boolean masks indicating the pixels containing the CME (black) according to ALMANAC. Note that the first and last frame are immediately before and after the event according to ALMANAC. The same information for CMEs 3, 11, and 16 are shown in Figures\,\ref{fig:example2}\,--\,\ref{fig:example4}.

In Figure\,\ref{fig:example1} the CME can be seen to erupt from the active region situated on the western limb in the SDO/AIA images (\textit{left}). The event is captured by ALMANAC as the CME propagates (\textit{right}). This is more clear in the animated version of this figure, which is available online.

Similarly, Figure\,\ref{fig:example2} shows the eruption from a disk central CME, which is associated with an M class flare between 01:18 and 03:00. This is again captured by ALMANAC, which may be viewed in the corresponding animated figure. As the CME appears to propagate radially from the source location, the majority of the Boolean mask is saturated from this event. The AIA instrument automatically adjusts exposure time in the presence of large flares in order to avoid detector saturation (for example, the exposure times can drop from 2\,s to 0.1\,s). Despite normalising the data counts by the exposure times, and limiting the intensities to a maximum threshold, this variation of the exposure time causes problems for our automated procedures. For example, the signal\,--\,to\,--\,noise ratio for the short-exposure observations will be reduced, which may contribute to this global mask saturation in ALMANAC. However, in the event of a CME associated with a major flare, ALMANAC's estimate of the source location or onset time remains stable (Table\,\ref{table:almanac}).

In Figures\,\ref{fig:example3} and \,\ref{fig:example4}, sigmoid and filament eruptions that have no flare association according to the CDAW catalogue are presented. These events are correctly detected by ALMANAC and the evolution of \cmed\ (Figure\,\ref{fig:example4}) may be viewed in the corresponding animated version.

From these four examples it is clear that ALMANAC is capable of detecting a variety of eruptive events on the solar disk using SDO/AIA 193\,\AA\ observations. The center of mass of each event, as defined in the method section, is indicated by the green asterisks in Figure\,\ref{fig:com}. These locations, along with CME onset time are compared to the CDAW values in Table\,\ref{table:almanac}.

Focusing firstly on the onset times of the twenty eruptions we see that ALMANAC detects the eruptions within absolute mean values of \fullatime\ of those reported in CDAW. For longitude and latitude, we find that ALMANAC absolute mean values are within \fulllongi\ and \fulllati\ of CDAW, respectively. However, these results are skewed by one outlying event comparison: there is a front\,-\,sided CME within the ALMANAC observational window which is listed as a far\,-\,side halo\,-\,CME in the CDAW catalogue (Table\,\ref{table:almanac}; CME 5). This event is discussed in more detail in \S\,\ref{sec:cme5}. Discarding this event, the onset times for ALMANAC are within \atime\ of CDAW whilst differences in longitude and latitude become \longi\ and \lati, respectively. This equates to the absolute mean distance of the CDAW and ALMANAC source regions being \dist. We note that CDAW typically records the source region using X\,--\,ray data, whilst ALMANAC uses EUV, which may lead to small differences.

\subsubsection{Investigating CME\,5}\label{sec:cme5}
\begin{figure*}
\centerline{\includegraphics[width=0.7\textwidth]{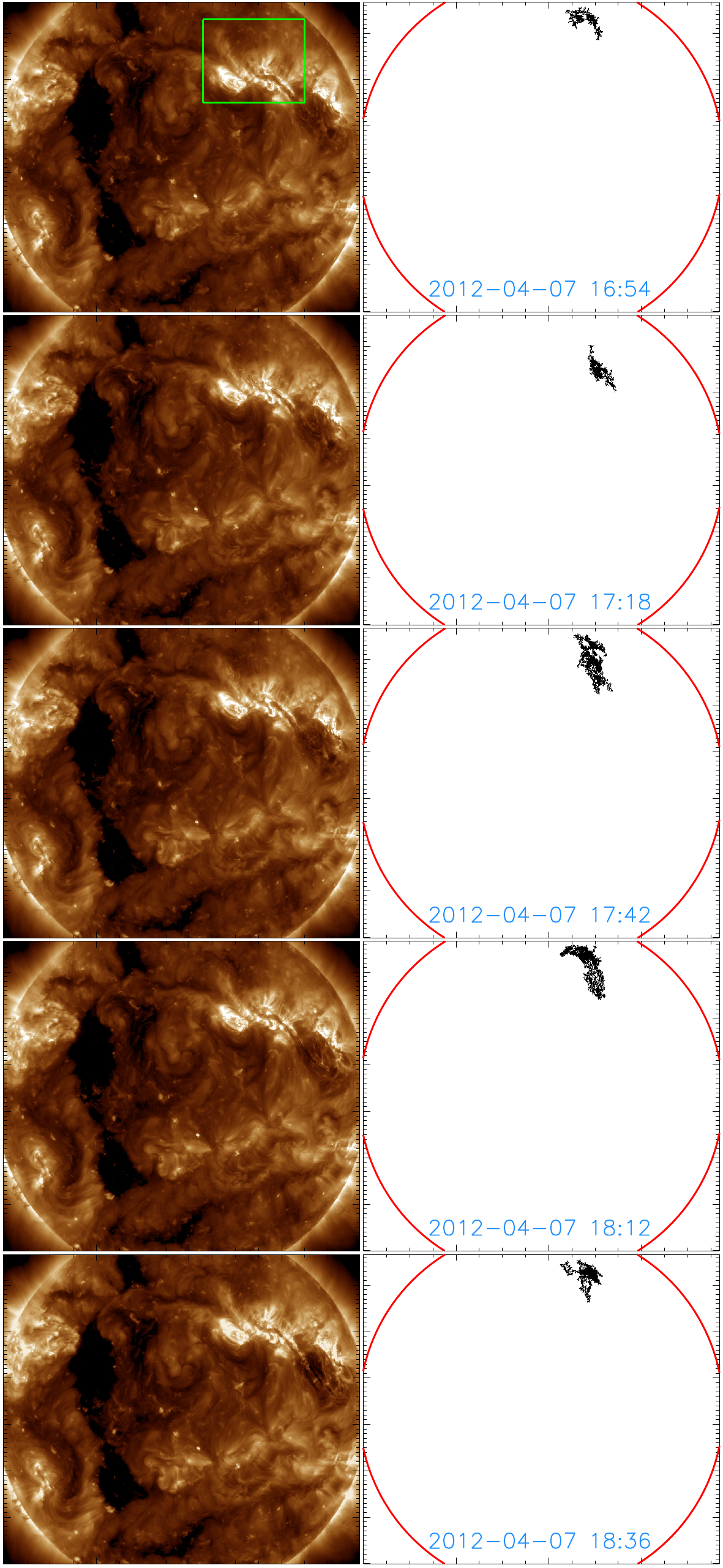}}
\caption{\textit{Left}: Synoptic SDO/AIA images taken at regular intervals (time increases in descending order) for CME\,5. The green box in the first image indicates the region from which the CME emerges. \textit{Right}: The corresponding Boolean mask generated by ALMANAC isolating the eruptive pixels (black). The solar limb is indicated in red. Note that the eruption is difficult to see in the SDO/AIA images but is evident in the animated version of this figure (online only).}
\label{fig:example5a}
\end{figure*}
\begin{figure*}
\centerline{\includegraphics[width=0.7\textwidth]{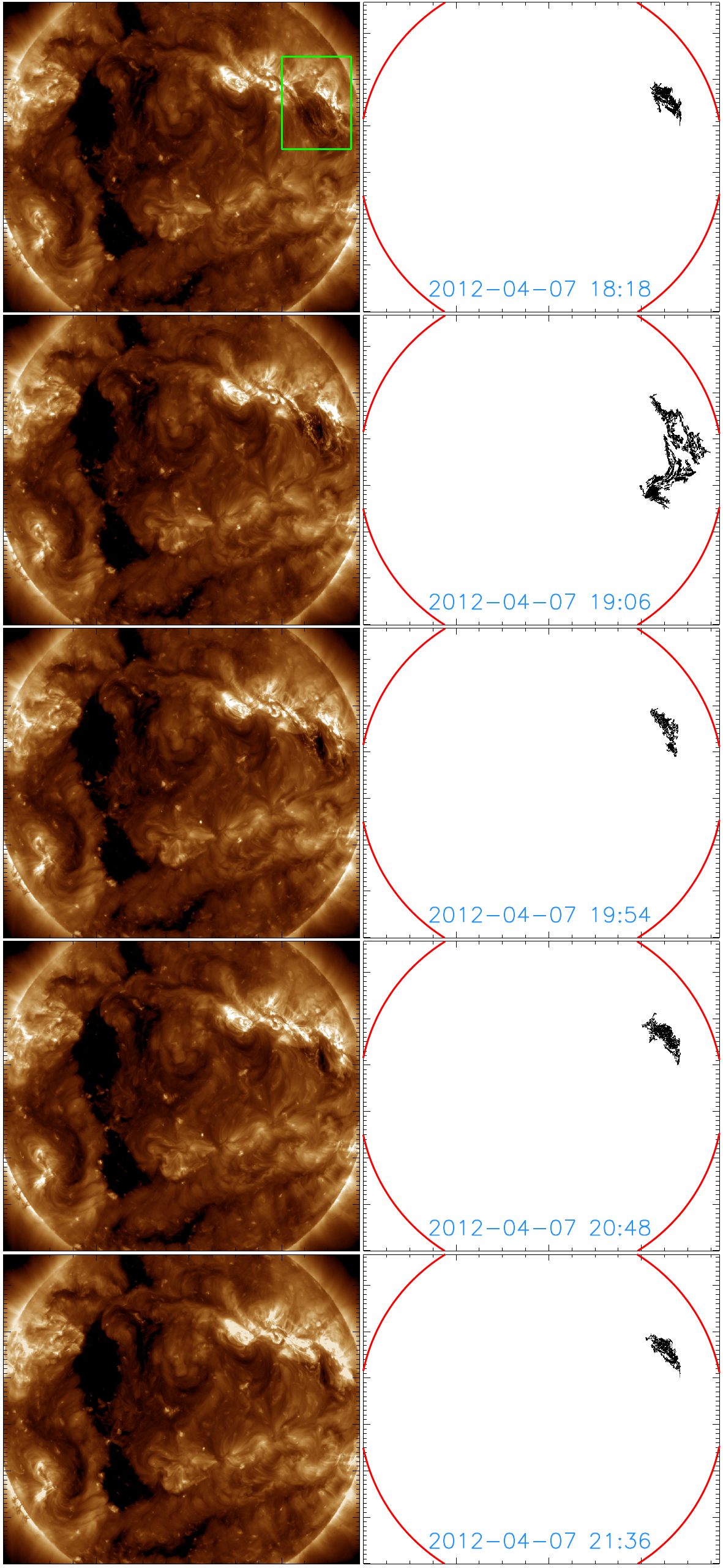}}
\caption{\textit{Left}: Synoptic SDO/AIA images taken at regular intervals (time increases in descending order) for CME\,5. The green box in the first image indicates the region from which the CME emerges. \textit{Right}: The corresponding Boolean mask generated by ALMANAC isolating the eruptive pixels (black). The solar limb is indicated in red. Note that the eruption is difficult to see in the SDO/AIA images but is evident in the animated version of this figure (online only).}
\label{fig:example5b}
\end{figure*}

In this subsection, CME\,5 is presented in Figures\,\ref{fig:example5a} and \ref{fig:example5b}. In this example\,--\,as is evident from the animated versions of these figures\,--\,a perturbation arises from a shearing loop arcade above a polarity inversion line. In ALMANAC, the north and south components of the CME is segmented into two different regions. The northerly propagating perturbation is seen to incite an oscillating disturbance in the filament before quelling, whilst the southerly component leads to a CME occurring at 18:18\,UT, S\,3\,W\,55. According to the CDAW halo\,-\,CME catalogue the source event is recorded on the far\,-\,side an hour later (and thus not detectable by ALMANAC using SDO/AIA data), although a manual inspection of the SDO/AIA data clearly shows an event on the near-side west limb. This highlights the need for caution in solely using ALMANAC for operational space weather. If a far\,-\,side halo\,-\,CME were to occur around a similar time to a near\,-\,side rapid temporal change, then ALMANAC would incorrectly associate the near\,-\,side event with the CME.


\section{Concluding Remarks}\label{sec:conc}
In this manuscript a novel approach to approximate the source location of halo\,-\,CMEs using SDO/AIA NRT data is presented. The method adopts similar approaches to the feature detection algorithms employed by coronal Moreton wave detection schemes \cite{long14,ireland19}. A total of twenty events from the CDAW halo\,-\,CME catalogue are identified for testing the new method, whereby it is found that our results are within \atime\ and \dist\ of the reported CDAW source region. Given the difficulty in accurately determining the source location of a CME with autonomous methods the approach developed in this manuscript performs well.

In Table\,\ref{table:almanac}, several halo CMEs are identified that are not associated with an X\,--\,ray source; these CMEs arise from filament/prominence eruptions. In the SDO/AIA 304 passband, filaments and prominences are distinctly visible against the surrounding atmosphere compared to other SDO/AIA channels. As such, it may be beneficial to optimise ALMANAC for use with the SDO/AIA 304 passband to further aid the detection of prominence eruptions. Furthermore, optimising the ALMANAC thresholds for detection for SDO/AIA passbands with sensitivity to higher temperatures such as 131, 335, or 94 may provide more accurate source regions for CMEs associated with large flares. Together with the SDO/AIA 193 passband, these additional channels would allow for multi-wavelength monitoring to supplement the results from the main ALMANAC detection passband and potentially increase sensitivity for events that are difficult to detect.

\citeA{murray18} demonstrate a back\,-\,propagation method using STEREO's Heliospheric Imagers \cite[HI]{hi} to compare the CME source location with the Solar Monitor Active Region Tracker \cite[SMART]{higgins11}, and flare forecasts. Similarly to ALMANAC, \citeA{murray18} find their method is typically within 5\,\textdegree\ of their reference catalogue, however, this back\,-\,propagation method is significantly more computationally expensive than the method developed here for operational space weather forecasting.

A requirement of ALMANAC is that it is efficient to run\,--\,on the order of seconds to minutes rather than hours to days. On a fairly modest Linux Mint system (Intel Core i7\,-\,9700 @\,3.00\,GHz, 16\,GB 2666\,MHz RAM with a 500\,GB Western Digital 5400\,RPM HDD) ALMANAC can process 8\,hours of data at a cadence of 6 minutes in $\approx$40\,s, which makes it suitable to be used as part of a real\,--\,time detection system.

\citeA{riley18} showed that current space weather forecasting models can typically predict the arrival times of CMEs to within $\pm10$\,hours, though the uncertainties surrounding this are large ($\pm20$\,hours). However, \citeA{barnard20} demonstrate that using weighted ensembles with HUXt that a mean reduction in arrival time error up to $20.1 \pm 4.1$\,\% and a mean reduction in arrival time uncertainty of $15.0 \pm 7.2$\,\% relative to unweighted ensembles are possible. It is also noted by \citeA{barnard20} that the method could be adopted in real\,-\,time if such observations for STEREO/HI existed. In a similar manner it is demonstrated by \citeA{barnard22} that an L5 observer could result in arrival time errors of approximately $6-8 \pm 1$\,hour for a sample of four CMEs. Our aim is to incorporate ALMANAC into an automated CME forecasting suite in order to provide similar improvements.

For the 160\,hours of synoptic AIA data analysed in this manuscript, ALMANAC records \detections\ events, whilst only twenty halo CMEs are identified in CDAW during this period. Thus ALMANAC, as well as correctly detecting the halo CMEs, also detects flaring and transient loops, plumes, and changes in active region morphology. Furthermore, ALMANAC can also  double-count some large events if their source regions become bifurcated (as discussed in \S\,\ref{sec:cme5}). The number of false\,-\,positive detections can be reduced by adjusting key thresholds in ALMANAC, and we have used only moderately stringent values in this work so that all potential events are detected. More stringent thresholds carry the risk of not detecting CMEs on the solar disk. As such, we will investigate refinements to ALMANAC to both reduce the number of false positives and to provide an ability to categorise events. Our immediate aim is to use ALMANAC as part of a comprehensive CME detection system, where large CMEs detected in coronagraph data can be associated with specific ALMANAC events, thus providing valuable constraints on the timing, locations, and early acceleration profiles of CMEs. Initially, the integration of ALMANAC with current detection suites will likely require a trigger from coronagraph detection codes. As it may take a few hours for a Sun\,-\,Earth directed halo\,-\,CME to become visible in coronagraph data, it is planned to incorporate an independent algorithm that will aim to score ALMANAC events on the likelihood of them being an Earth directed halo CME. If successful, then in addition to providing a source location for potential CMEs, ALMANAC could also provide an early\,-\,warning system for space weather events.

CMEs may often undergo deflection in the low corona \cite[for example]{majumdar20} and so one may not assume a CME propagates in the direction of its source. Considering this, combining the results from ALMANAC with similar results from coronagraph detection methods being developed as part of SWEEP could allow rudimentary triangulation to be performed to estimate the potential trajectory and velocity of the CME. This information would then provide observational constraint on the ensembles employed by CME propagation models such as HUXt \cite{owens20}.

\section{Open Research}
This work utilises the publicly available synoptic data from SDO/AIA \cite{sdo,aia}, which was processed using the Interactive Data Language (IDL) version 8.4 and the SolarSoft library. The source code and SDO/AIA data used for this manuscript can be obtained at \url{https://zenodo.org/badge/latestdoi/555290146}.

\acknowledgments
TW and HM gratefully acknowledge support by Leverhulme grant RPG-2019-361. HM also acknowledges STFC grant ST/S000518/1.
\clearpage

\bibliography{ref} 

%
%
\clearpage
\end{document}


%
%


\title{Supporting Information for ``Automated detection of coronaL MAss ejecta origiNs for space weather AppliCations (ALMANAC)"}


%
%

\authors{Thomas Williams\affil{1}, Huw Morgan\affil{1}}
\affiliation{1}{Department of Physics, Aberystwyth University, Penglais, 
Aberystwyth, SY23 3BZ, UK}

%
%

%

\begin{article}

%
%

\noindent\textbf{Contents of this file}
\noindent\textbf{Additional Supporting Information (Files uploaded separately)}
\begin{enumerate}
\item Captions for Movies S1 to S6
\end{enumerate}

\noindent\textbf{Introduction}
In this supporting document are the captions for the animated versions of the figures to be included with the online version of the manuscript.
\clearpage

\noindent\textbf{Movie S1.} 
Synoptic SDO/AIA 193 image sequence (\textit{left}) and the corresponding ALMANAC mask (\textit{right}) for \cmea. In the Boolean mask generated by ALMANAC, the pixels associated with the CME are indicated in black and the solar limb is indicated in red. This is a supplement to Figure\,\ref{fig:example1}.

\noindent\textbf{Movie S2.} 
Synoptic SDO/AIA 193 image sequence (\textit{left}) and the corresponding ALMANAC mask (\textit{right}) for \cmeb. In the Boolean mask generated by ALMANAC, the pixels associated with the CME are indicated in black and the solar limb is indicated in red. This is a supplement to Figure\,\ref{fig:example2}.

\noindent\textbf{Movie S3.} 
Synoptic SDO/AIA 193 image sequence (\textit{left}) and the corresponding ALMANAC mask (\textit{right}) for \cmec. In the Boolean mask generated by ALMANAC, the pixels associated with the CME are indicated in black and the solar limb is indicated in red. This is a supplement to Figure\,\ref{fig:example3}.

\noindent\textbf{Movie S4.} 
Synoptic SDO/AIA 193 image sequence (\textit{left}) and the corresponding ALMANAC mask (\textit{right}) for \cmed. In the Boolean mask generated by ALMANAC, the pixels associated with the CME are indicated in black and the solar limb is indicated in red. This is a supplement to Figure\,\ref{fig:example4}.

\noindent\textbf{Movie S5.} 
Synoptic SDO/AIA 193 image sequence (\textit{left}) and the corresponding ALMANAC mask (\textit{right}) for CME\,5. In the Boolean mask generated by ALMANAC, the pixels associated with the CME are indicated in black and the solar limb is indicated in red. This is a supplement to Figure\,\ref{fig:example5a}.

\noindent\textbf{Movie S6.} 
Synoptic SDO/AIA 193 image sequence (\textit{left}) and the corresponding ALMANAC mask (\textit{right}) for CME\,5. In the Boolean mask generated by ALMANAC, the pixels associated with the CME are indicated in black and the solar limb is indicated in red. This is a supplement to Figure\,\ref{fig:example5b}.

%
%
\end{article}
\clearpage